\documentclass[sigconf]{acmart}
\usepackage[per-mode=symbol,per-symbol=p]{siunitx}
\usepackage[utf8]{inputenc}
\usepackage{adjustbox}
\usepackage[ruled,vlined]{algorithm2e}
\usepackage{algorithmic}
\usepackage{multirow}
\usepackage{graphicx}
\usepackage{textcomp}
\usepackage{xcolor}
\usepackage{balance}
\usepackage{enumitem}
\usepackage{setspace}
\usepackage{wrapfig}
\usepackage{listings}
\usepackage{color}
\usepackage{caption}
\usepackage{subcaption}
\usepackage{tikz}
\usetikzlibrary{shapes.geometric, arrows}
\usetikzlibrary{fit}
\usetikzlibrary{calc}
\tikzstyle{process_g} = [rectangle, minimum width=2.3cm, minimum height=0.75cm, text centered, text width=2.3cm, draw=black, fill=gray!40]

\tikzstyle{process} = [rectangle, minimum width=2.3cm, minimum height=0.75cm, text centered, text width=2.3cm, draw=black, fill=gray!10]
\tikzstyle{arrow} = [thick,->,>=stealth]

\newcommand{\etal}{\emph{et al.}\xspace}
\newcommand{\ie}{\emph{i.e.}, }
\newcommand{\eg}{\emph{e.g.}, }

\newcommand{\VVC}{\emph{Versatile Video Coding }}

\renewcommand{\footnotesize}{\fontsize{7pt}{7pt}\selectfont}
\copyrightyear{2024}
\acmYear{2024}
\setcopyright{rightsretained}
\acmConference[MHV '24]{Mile-High Video Conference}{February 11--14, 2024}{Denver, CO, USA}
\acmBooktitle{Mile-High Video Conference (MHV '24), February 11--14, 2024, Denver, CO, USA}
\acmDOI{10.1145/3638036.3640805}
\acmISBN{979-8-4007-0493-2/24/02}

\begin{document}

\title{Gain of Grain:  A Film Grain Handling Toolchain for VVC-based Open Implementations}

\author{Vignesh V Menon}
\email{vignesh.menon@hhi.fraunhofer.de}
\orcid{0000-0003-1454-6146}
\affiliation{
  \institution{\small{Video Communication and Applications Dept.}}
  \institution{Fraunhofer HHI}
  \city{Berlin}
  \country{Germany}
}

\author{Adam Wieckowski}
\email{adam.wieckowski@hhi.fraunhofer.de}
\affiliation{
  \institution{\small{Video Communication and Applications Dept.}}
  \institution{Fraunhofer HHI}
  \city{Berlin}
  \country{Germany}
}

\author{Jens Brandenburg}
\email{jens.brandenburg@hhi.fraunhofer.de}
\affiliation{
  \institution{\small{Video Communication and Applications Dept.}}
  \institution{Fraunhofer HHI}
  \city{Berlin}
  \country{Germany}
}

\author{Benjamin Bross}
\email{benjamin.bross@hhi.fraunhofer.de}
\affiliation{
  \institution{\small{Video Communication and Applications Dept.}}
  \institution{Fraunhofer HHI}
  \city{Berlin}
  \country{Germany}
}

\author{Thomas Schierl}
\email{thomas.schierl@hhi.fraunhofer.de}
\affiliation{
  \institution{\small{Video Communication and Applications Dept.}}
  \institution{Fraunhofer HHI}
  \city{Berlin}
  \country{Germany}
}

\author{Detlev Marpe}
\email{detlev.marpe@hhi.fraunhofer.de}
\affiliation{
  \institution{\small{Video Communication and Applications Dept.}}
  \institution{Fraunhofer HHI}
  \city{Berlin}
  \country{Germany}
}

\renewcommand{\shortauthors}{Vignesh V Menon~\etal}

\begin{abstract}
Film grain is a distinctive visual characteristic cherished by filmmakers and cinephiles for its ability to evoke nostalgia and artistic aesthetics. However, faithful preservation of film grain during encoding poses unique challenges. Film grain introduces random noise, complicating traditional compression techniques. Consequently, specialized algorithms and encoding strategies have emerged, aiming to strike a harmonious equilibrium. This paper delves into the nuanced realm of film grain handling in \VVC~(VVC) encoding. We explore the delicate balance between retaining the cinematic charm of film grain and achieving efficient compression. 
Moreover, we discuss the importance of perceptual quality assessment and adaptive encoding techniques in preserving film grain fidelity. Additionally, we delve into the impact of film grain handling on bitrate control and compression efficiency using VVenC, an open and optimized VVC encoder. Understanding the role of film grain and its nuanced treatment within encoders becomes increasingly pivotal for delivering high-quality, grain-inclusive content in the digital age.
\end{abstract}




\maketitle


\section{Introduction}
Film grain, an inherent characteristic of analog film, contributes to the unique visual aesthetics and cinematic experience in movies~\cite{fg_ref5}. With the evolution from analog to digital video formats, film grain preservation and coding have gained attention for their crucial role in maintaining the intended cinematic look and feel. With the advent of digital video, the need arose to adapt the digital medium to emulate film's organic and textured appearance, leading to film grain emulation techniques. In video encoding, preserving film grain is crucial for maintaining the intended aesthetics of movies initially shot on film. Filmmakers and studios want to ensure that digital versions of their films retain the look and feel of the original celluloid, including its grain structure. Film grain handling becomes essential in restoring classic films or archival content. Techniques for preserving or reintroducing the original film grain allow the faithful preservation of historical works. 

\begin{figure}[t]
\centering
\begin{subfigure}{0.235\textwidth}
    \centering
    \includegraphics[width=\textwidth]{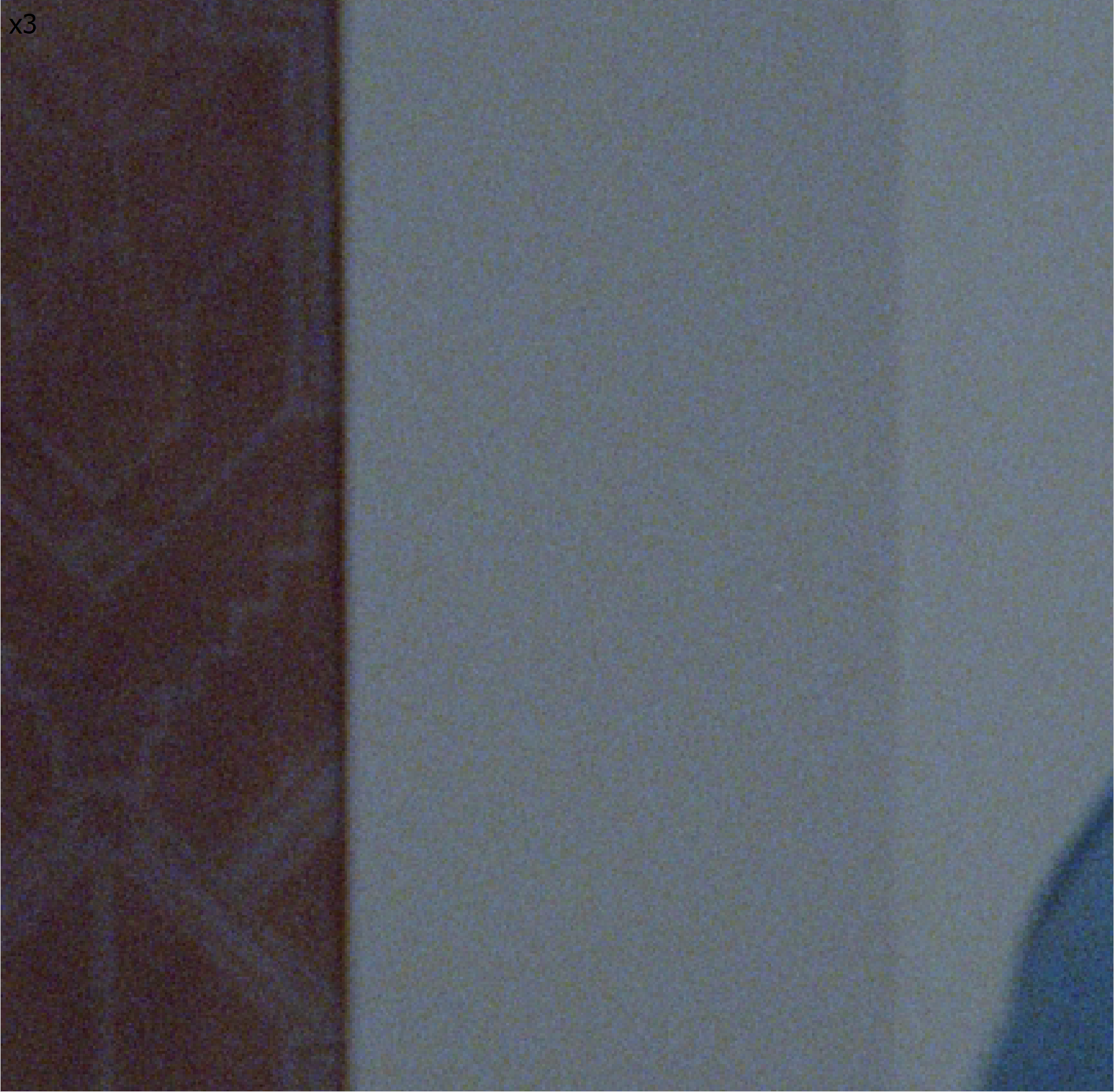}
    \caption{original}
\end{subfigure}
\hfill
\begin{subfigure}{0.235\textwidth}
    \centering
    \includegraphics[width=\textwidth]{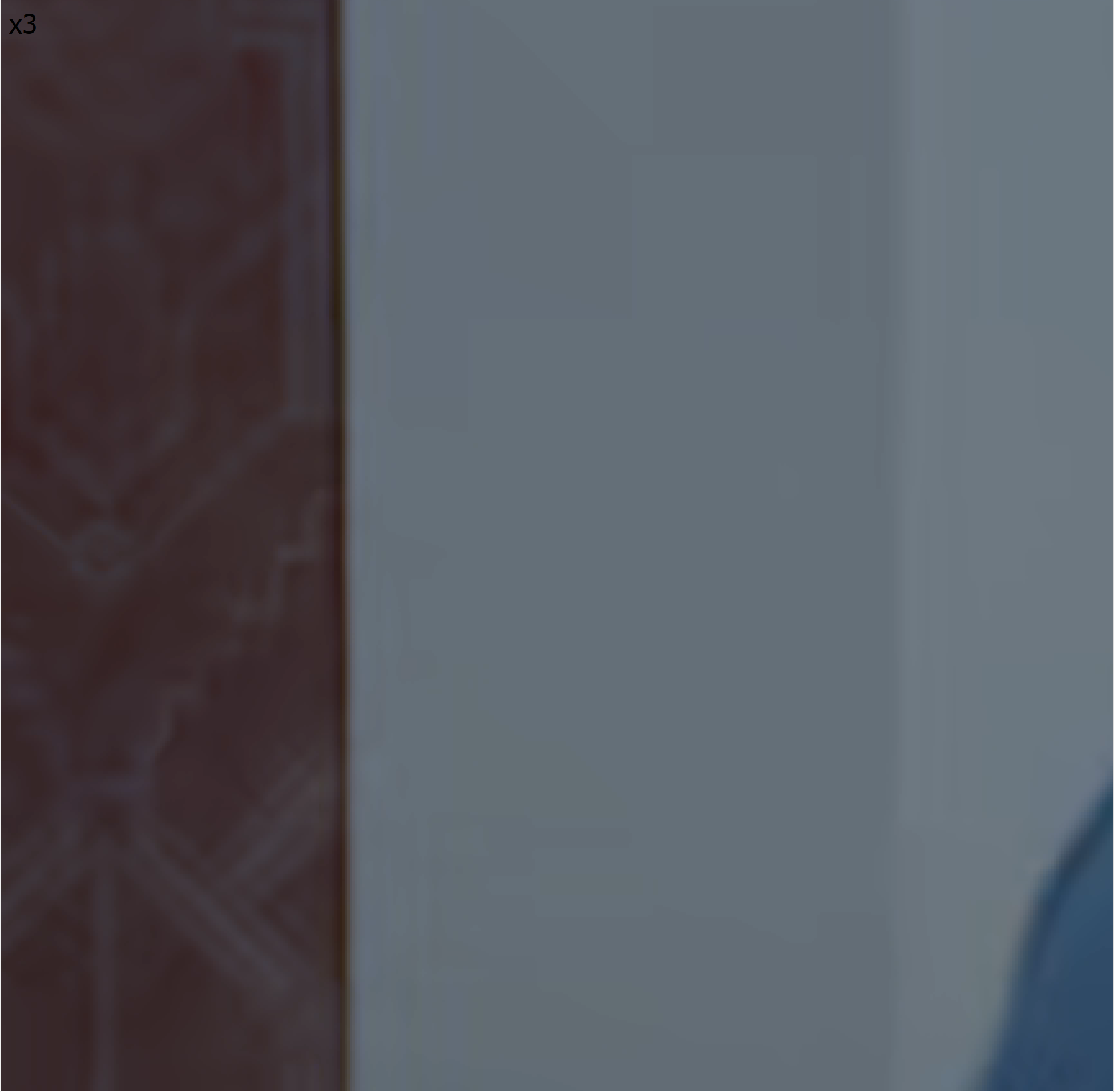}
    \caption{encoded at 600 kbps}
\end{subfigure}
\caption{Illustration of compression artifacts introduced due to film grain when encoded using VVenC encoder at low bitrates.}
\label{fig:ghost_intro_plot}
\end{figure}

\begin{figure*}[t]
\centering
    \includegraphics[width=\textwidth]{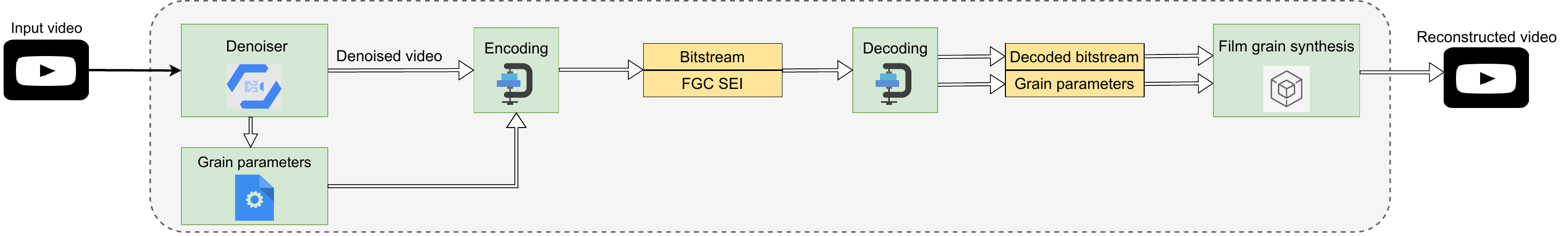}
\caption{State-of-the-art film grain handling toolchain framework.}
\label{fig:sota_arch}
\end{figure*}

The emergence of the \VVC~(VVC)~\cite{vvc_ref, vvenc_toolchain} standard brings new opportunities and challenges in efficiently representing film grain, aiming to preserve its artistic value while ensuring compatibility with modern video compression techniques. 
Figure~\ref{fig:ghost_intro_plot} illustrates the blockiness and other compression artifacts introduced when grainy video content is encoded at low bitrates. Film grain presents a challenge in video coding due to its random and non-uniform nature, which can amplify compression artifacts if not adequately handled during encoding. Coding film grain in VVC applications demands efficient techniques that balance preserving the cinematic texture and optimizing compression efficiency. Moreover, as high-resolution and high-dynamic-range (HDR) content becomes more prevalent, the efficient representation and faithful reproduction of film grain in VVC-encoded video streams are essential for delivering superior visual quality and maintaining artistic intent. Moreover, integrating film grain handling into VVC requires adherence to standardized encoding practices while accommodating unique film grain characteristics~\cite{Ameur2022DeepBasedFG}. It is expected to address a delicate interplay of artistic, perceptual, and technical factors~\cite{fg_ref5,fg_ref4}. In many state-of-the-art film grain handling methods, film grain is filtered from the original video sequence as a pre-processing step before encoding and synthesized as a post-processing step after decoding~\cite{gomila_ref, fg_ref1, fg_ref3}.

This paper presents a \textit{comprehensive compression efficiency, time complexity, and qualitative analysis} of a film grain handling toolchain using state-of-the-art VVC-based open-source video coding. This highlights the complexity of this relationship and the need for adaptive solutions that balance video quality and energy efficiency.

\section{Background}
Film grain introduces challenges in video encoding. While high-efficiency video codecs like VVC aim to compress video content as efficiently as possible, the random nature of grain can hinder compression algorithms. Encoding can lead to grain loss if not handled carefully, impacting the film's authenticity. Film grain handling in video encoding involves various techniques, including adaptive encoding algorithms that can identify grainy regions and apply different compression settings to preserve the grain. Additionally, post-processing filters can be applied to reintroduce grain in digitally shot content. Managing film grain during encoding requires a careful balance between preserving quality, retaining aesthetics, and optimizing bitrate for streaming or storage. This trade-off is a critical consideration in video encoding.

\subsection{Related work}
The film grain noise modeling technique for video coding was initially proposed by Gomila~\etal~\cite{gomila_ref}, introducing noise removal as an initial phase in video encoding, followed by noise synthesis as a subsequent step in video decoding. This concept received standardization acknowledgment from AVC~\cite{avc_grain_ref}. Ozkan~\etal~\cite{denoise_ref1} explored temporal filtering techniques aimed at noise suppression while preserving edge details. Concurrently, integrated spatial and temporal filtering strategies were examined in studies such as~\cite{denoise_ref2,denoise_ref3}, employing block motion estimation for temporal filtering. Boo~\etal~\cite{denoise_boo} implemented the Karhunen–Loeve (KL) transform along the temporal axis to decorrelate frame dependencies, coupled with adaptive Wiener filtering for frame smoothing. In the denoising process, finding an edge region of the image is essential since most denoising algorithms tend to blur the image, especially around the edges. Canny~\cite{canny} proposed an edge detection algorithm for noisy images.

Two different methods were proposed for film grain synthesis. One is to use the film grain database for low complexity. The film grain pattern is first identified, and the decoder generates a larger film grain size from a smaller film grain stock. However, the block-based copy-and-paste method might yield artificial boundary artifacts. Besides, the method is workable only when the film stock information is known a priori. The other is to use some models for blind film grain synthesis. Several methods have been proposed, \eg high-order statistics based~\cite{fgs_ref1,fgs_ref2}, parametric-modeling-based~\cite{fgs_ref3,fgs_ref4} or patch-based noise synthesis methods~\cite{fgs_ref5}.

Radosavljevic~\etal~\cite{vvc_grain_ref} proposed a complete VVC software implementation of the film grain feature, including film grain analysis and on-the-fly parameter estimation at the encoder side and film grain synthesis module at the decoder side. Many other works in the literature describe film grain characterization and film grain synthesis. In addition, the specification presented in~\cite{smpte_rdd5_ref}, also known as the SMPTE-RDD5 model, describes the film grain characterization and synthesis approach in the frequency domain. It precisely defines the synthesis part and specifies a bit-accurate grain blending process.

\begin{figure*}[t]
\centering
    \includegraphics[width=\linewidth]{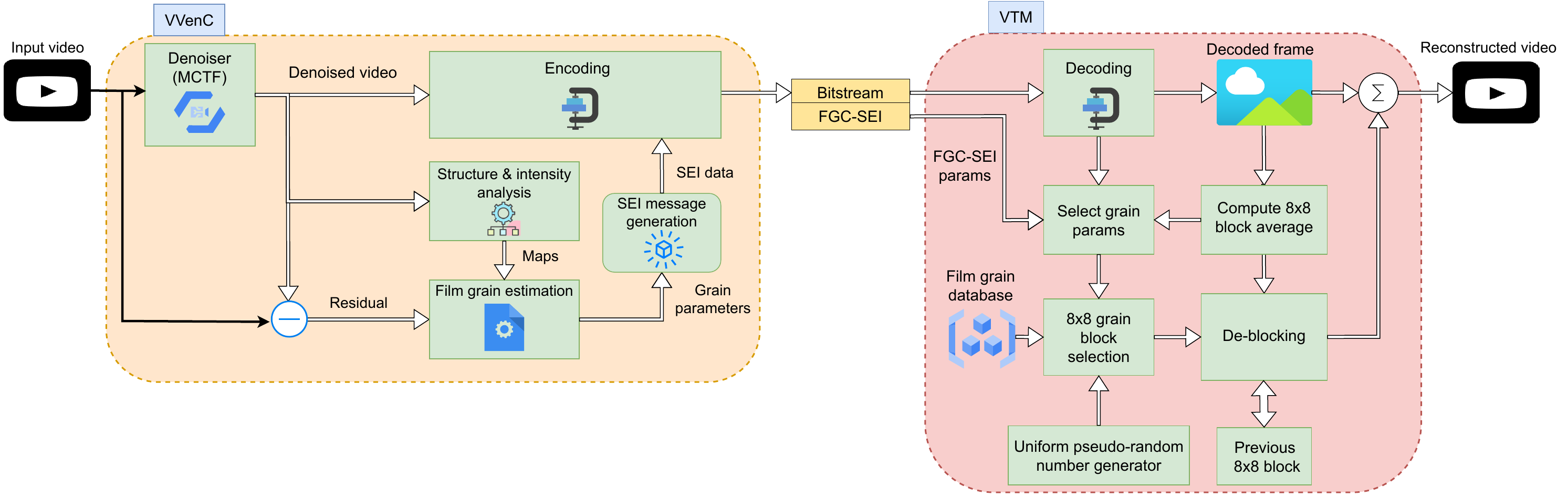}
\caption{Film grain handling toolchain considered in this paper.}
\label{fig:grain_arch}
\end{figure*}

\subsection{State-of-the-art}
The architecture of the state-of-the-art film grain handling toolchain is shown in Figure~\ref{fig:sota_arch}, according to which it is classified into four steps, as described below. 
\paragraph{Denoising} involves reducing or removing noise from the video, including film grain. It is often performed as a pre-processing step to enhance visual quality and reduce artifacts. It improves the perceptual quality of the video by reducing unwanted noise, including film grain, before encoding. The extracted residual noise is used to model the film grain characteristics, such as spatial distribution, intensity variations, and temporal behavior.

\paragraph{Encoding} involves compressing the video data for efficient storage and transmission. Film grain characteristics must be considered to balance compression efficiency and quality preservation. It achieves high compression ratios while maintaining the essential characteristics of film grain.

\paragraph{Decoding} reconstructs the video data from the compressed format. To ensure visual fidelity, the film grain representation must be faithfully reproduced during decoding. It reconstructs the video with accurate film grain representation for playback or further processing.

\paragraph{Grain synthesis} involves generating or enhancing film grain in the video. This can be applied during post-processing or restoration to recreate the aesthetic qualities of the original film. It preserves or introduces film grain to maintain the original content's visual characteristics or achieve a specific artistic look.

These processes are interconnected and collectively contribute to the overall handling of film grain in video coding. Denoising is often applied before encoding to improve the quality of the input video. During encoding, strategies are employed to represent film grain characteristics efficiently. Decoding aims to reconstruct the original film grain during playback. Additionally, grain synthesis can enhance or recreate film grain in the final output.


\section{Proposed toolchain implementation}
The film grain characteristics (FGC) supplementary enhancement information (SEI), as specified in AVC or the Versatile SEI (VSEI) standard for VVC, only provides the syntax to transmit model parameters to the decoder~\cite{fgc_sei_spec_itu,fgc_sei_spec_iso}. It neither provides the methods to estimate parameters nor how to synthesize film grain. It allows the encoder to use the FGC SEI message to characterize the film grain present in the source video and have film grain removed by pre-processing filtering and/or lossy compression. In the film grain analysis (FGA) of the proposed toolchain, motion compensated temporal filter (MCTF)~\cite{mctf_ref} used in VVenC filters out the film grain. Then, the denoised video is encoded. The film grain is known to have high levels at high frequencies (for example, in the DCT domain), which is usually suppressed by the quantization process. At the same time, film grain parameters are inserted in the FGC SEI messages. The decoder decodes the bitstream as well as the FGC SEI messages. It generates the decoded video, enhancing the film grain synthesis (FGS) process. Notably, FGS is optional and can be skipped, producing only decoded video without adding film grain.

\subsection{Denoising}
\label{sec:denoise}
Firstly, denoising is applied to the video frames to reduce the noise that the grain introduces. This step is crucial to balance the grain's aesthetic appeal and the need for clarity in the encoded video~\cite{fg_ref1}. The VVenC encoder~\cite{vvenc_ref} already employs a denoising stage~\cite{mctf_ref}, based on a framework proposed initially in \cite{mctf_ref_org}-- motion compensated temporal (pre-)filtering. The assumption behind the function of the filter is that the video content will be visible in multiple frames, while the noise will be different in each video frame. The filter performs a blockwise motion search for each filtered frame in neighboring frames to remove the noise. Using up to eight predictors, a weighted average
of the current frame block and its predictors is generated and used for further encoding. In~\cite{mctf_ref}, improved search strategies, reference number reduction, and flexible block size were introduced, improving the filter runtime and operation. The temporal pre-filter has already proven very efficient for denoising in video compression tasks and is thus the natural first choice for use in the film grain estimation framework using VVenC.

\subsection{Film grain estimation}
In the proposed toolchain, a framework for film grain handling based on frequency filtering and parameterization of grain is implemented in VVenC~\cite{vvenc_ref}, as shown in Figure~\ref{fig:grain_arch}. The pixel values of the denoised frames are subtracted from the original input frames to get the residual frame, which is used to estimate film grain. Similar to the approach described in ~\cite{vvc_grain_ref}, a film grain pattern is characterized using a horizontal high cut-off frequency and a vertical high cut-off frequency obtained in the discrete cosine transform domain. These characterize the grain size, spatial distribution, intensity variation, and grain structure. After obtaining the film grain pattern, it is scaled to the appropriate level using a stepwise scaling function that considers the underlying image's characteristics. Afterward, the film grain pattern is blended into the image using additive blending~\cite{sean2021fixed}. This estimation is essential for accurately representing the grain during encoding. To convey the grain parameters to the decoder, the encoder embeds it as supplemental enhancement information (SEI) in the bitstream~\cite{smpte_rdd5_ref}, as the "Film Grain SEI." The Film Grain SEI inherits the same syntax and semantics of the AVC film grain SEI message~\cite{avc_grain_ref}. 

Since we implement a frequency filtering model for film grain estimation, \emph{film\_grain\_model\_id} is set to 0. Additive blending~\cite{sean2021fixed} is used when \emph{blending\_mode\_id} is set to 0. Since our implementation analyses film grain for only the luma channel, \emph{comp\_model\_present\_ flag[0]} is set to 1. FGC SEI message is inserted at each frame, which is indicated by setting the \emph{film\_grain\_characteristics\_persistence\_flag} to 0. It also means the FGC SEI message only applies to the current decoded frame.

\subsection{Film grain synthesis}
The proposed toolchain utilizes the film grain synthesis implemented in the VTM reference software~\cite{vvc_grain_ref} illustrated in Figure~\ref{fig:grain_arch}. However, it is described in this paper to make it self-contained.

The process starts with creating a film grain pattern (block of $64\times64$ pixels) for all pairs of cut-off frequencies. \mbox{SMPTE-RDD5} defines a pre-computed set of transformed pseudo-random numbers, defined apriori (within \mbox{SMPTE-RDD5}), and stored for further use. The pre-computed set is obtained using an integer approximation of floating-point DCT, also defined within the specification. Thereafter, a $64\times64$ block of transformed pseudo-random values, denoted as $B$, undergoes a low-pass filtering. Each film grain pattern is synthesized using a different pair of cut-off frequencies. Therefore, horizontal high cutoff frequency and vertical high cut-off frequency define film
grain pattern. Low pass filtering is performed by setting to zero all coefficients of a block $B$ in such a way that $x>$horizontal cutoff or $y>$vertical cutoff leads to B[x; y] = 0, where x = \{0,...,  63 \} and y = \{0,...,  63 \} are horizontal and vertical coordinates within the block. After filtering, inverse DCT is performed to obtain block $b'$. By this, $b'$ represents the FG pattern. Different film grain patterns (for different cut-off pairs) can be pre-computed, generating a database of all available film grain blocks/patterns. The database includes $h\times v\times64\times64$ film grain samples (given the range of cut-off frequencies, it leads to $13\times13\times64\times64$). 

A film grain simulation is performed after the film grain database is created and after receiving an FGC SEI message. To choose a particular pattern from the film grain database, one can take advantage of the $8\times 8$ block average and the interval to which the average value of the currently processed $8\times 8$ block belongs (the currently processed block is the block to which we add a film grain is taken from the image we are processing - usually a decoded frame). By comparing the average value with the SEI message \emph{intensity\_interval\_lower\_bound[c][i]} and \emph{intensity\_interval\_upper\_bound [c][i]} parameters, the intensity interval is identified. Hence, based on the average value of the block and intensity intervals received with the FGC SEI, a selection of FGC parameters is performed. A selection includes the scaling parameter (\emph{comp\_model\_value[c][i][0]} ) and cut-off frequencies (\emph{comp\_model\_value[c][i][1]} and \emph{comp\_model\_value [c][i][2]} ). Selected cut-off frequencies are used to access the film grain database. Thereafter, film grain is added to the image on an 8x8 basis. Thus, 8x8 blocks are randomly selected from the FG block (size $64\times64$) created in the previous step. The pseudo-random number generator defines an offset from the origin of the $64\times64$ film grain block to ensure bit-exact simulation.

Additional scaling based on the scaling parameter conveyed in \emph{comp\_model\_value[c][i][0]} can be performed to get the appropriate intensity of the film grain. Deblocking is performed as well to smooth the edges. Finally, an input image can be processed block by block using $8\times8$ granularity in raster scan order or in any other convenient way.

\section{Evaluation}
This section illustrates our experimental setup to assess the performance of the proposed toolchain. It is compared to the default coding scheme regarding rate-distortion performance and energy consumption.

\subsection{Experimental setup}
We run experiments on an AMD EYPC 7502P processor (32 cores), where we run each VVenC v1.10 instance using four CPU threads, enabling adaptive quantization. All sequences are down-scaled to 1920x1080 8bit for evaluation. 
Table~\ref{tab:exp_par} summarizes the list of experimental parameters. We run two-pass rate control. We consider the \texttt{Default} toolchain as the benchmark, where we encode the input video using VVenC with MCTF disabled. We decode the resulting bitstream using the VTM decoder. Notably, FGA and FGS are disabled in this toolchain.

\begin{table}[t]
\caption{Experimental parameters used to evaluate the film grain handling toolchain.}
\centering
\resizebox{\columnwidth}{!}{
\begin{tabular}{l|c||c}
\specialrule{.12em}{.05em}{.05em}
\specialrule{.12em}{.05em}{.05em}
\emph{Parameter} & \emph{Notation} & \emph{Values}\\
\specialrule{.12em}{.05em}{.05em}
\specialrule{.12em}{.05em}{.05em}
Resolution height [pixels] & $r$ &  1080   \\
\hline
Set of bitrates & $\mathcal{B}$ & \{ 0.25, 0.50, 1.00, 2.00, 3.00, 4.00, 6.00 \}  \\
\hline
Set of presets [VVenC] & $\mathcal{P}$ & \{ faster, medium, slower \}  \\
\hline
\multicolumn{2}{c||}{Configuration} & random access  \\
\hline
\multicolumn{2}{c||}{Quality metrics} &  MS-SSIM YUV, PSNR YUV  \\
\specialrule{.12em}{.05em}{.05em}
\specialrule{.12em}{.05em}{.05em}
\end{tabular}
}
\label{tab:exp_par}
\end{table}
\begin{figure}[t]
\centering
\begin{subfigure}{0.32\columnwidth}
    \centering
    \includegraphics[width=\textwidth]{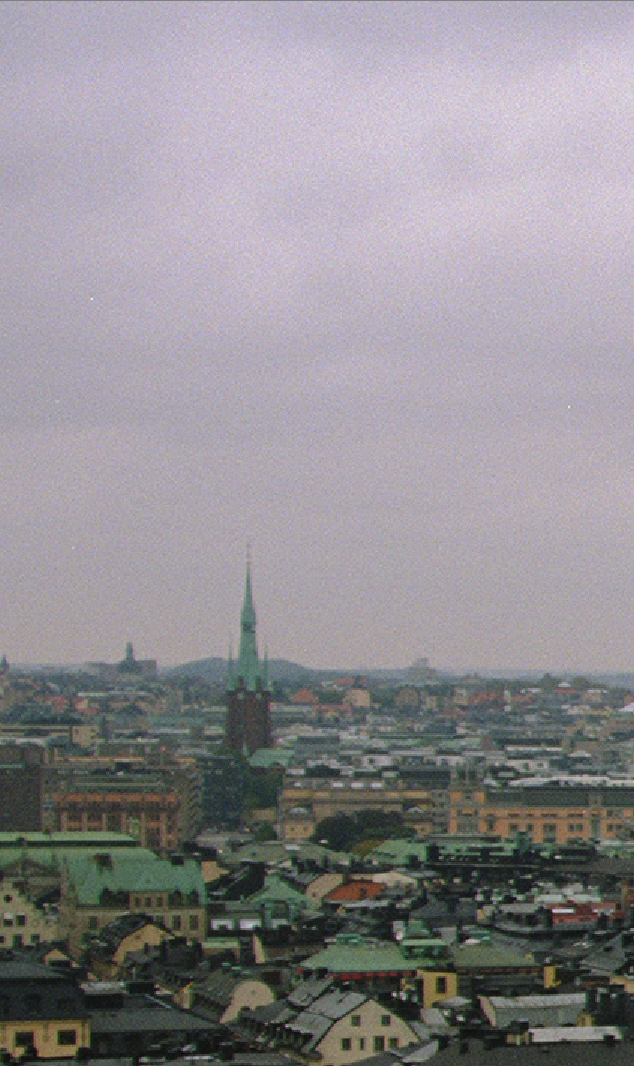}
    \caption{}
\end{subfigure}
\hfill
\begin{subfigure}{0.32\columnwidth}
    \centering
    \includegraphics[width=\textwidth]{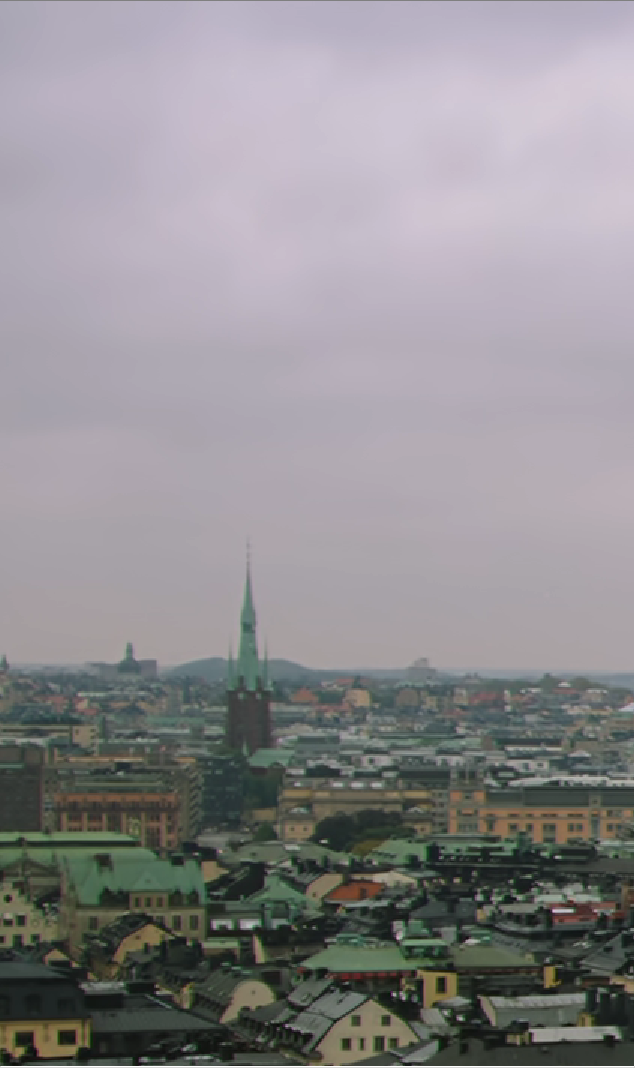}
    \caption{}
\end{subfigure}
\hfill
\begin{subfigure}{0.32\columnwidth}
    \centering
    \includegraphics[width=\textwidth]{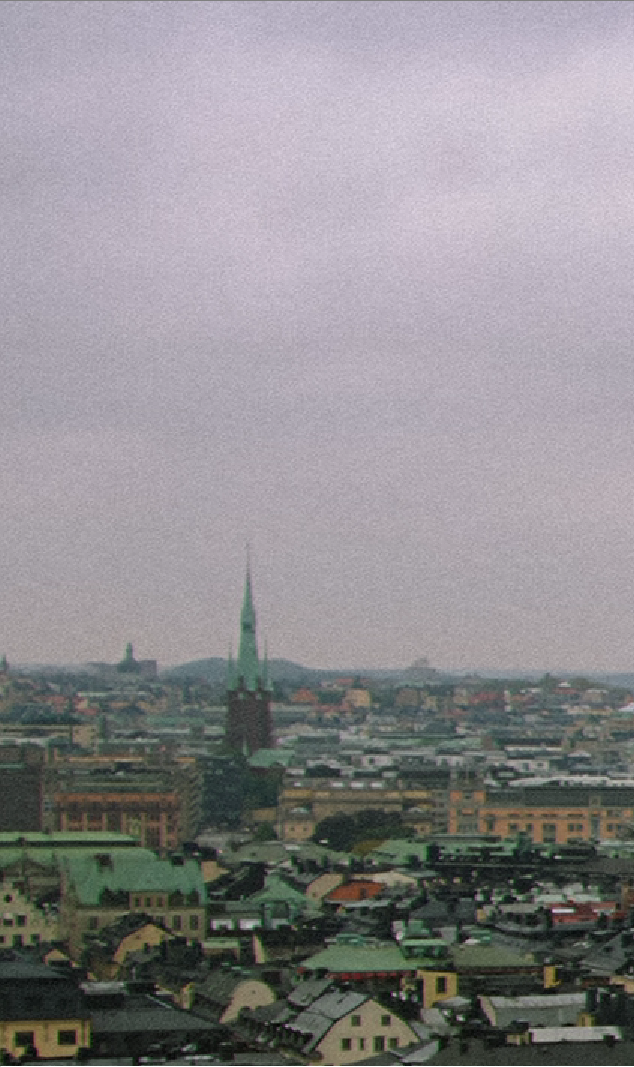}
    \caption{}
\end{subfigure}
\vspace{-0.3em}
\caption{Comparison of (a) original sequence, (b) encoded at 3 Mbps with MCTF, and (c) encoded at 3 Mbps with the proposed toolchain (with FGS) of the representative (cropped) frame of \textit{OldTownCross} sequence.}
\label{fig:oldtowncross_plot}
\end{figure}

\begin{figure*}[t]
\centering
\begin{subfigure}{0.45\textwidth}
    \centering
    \includegraphics[width=\textwidth]{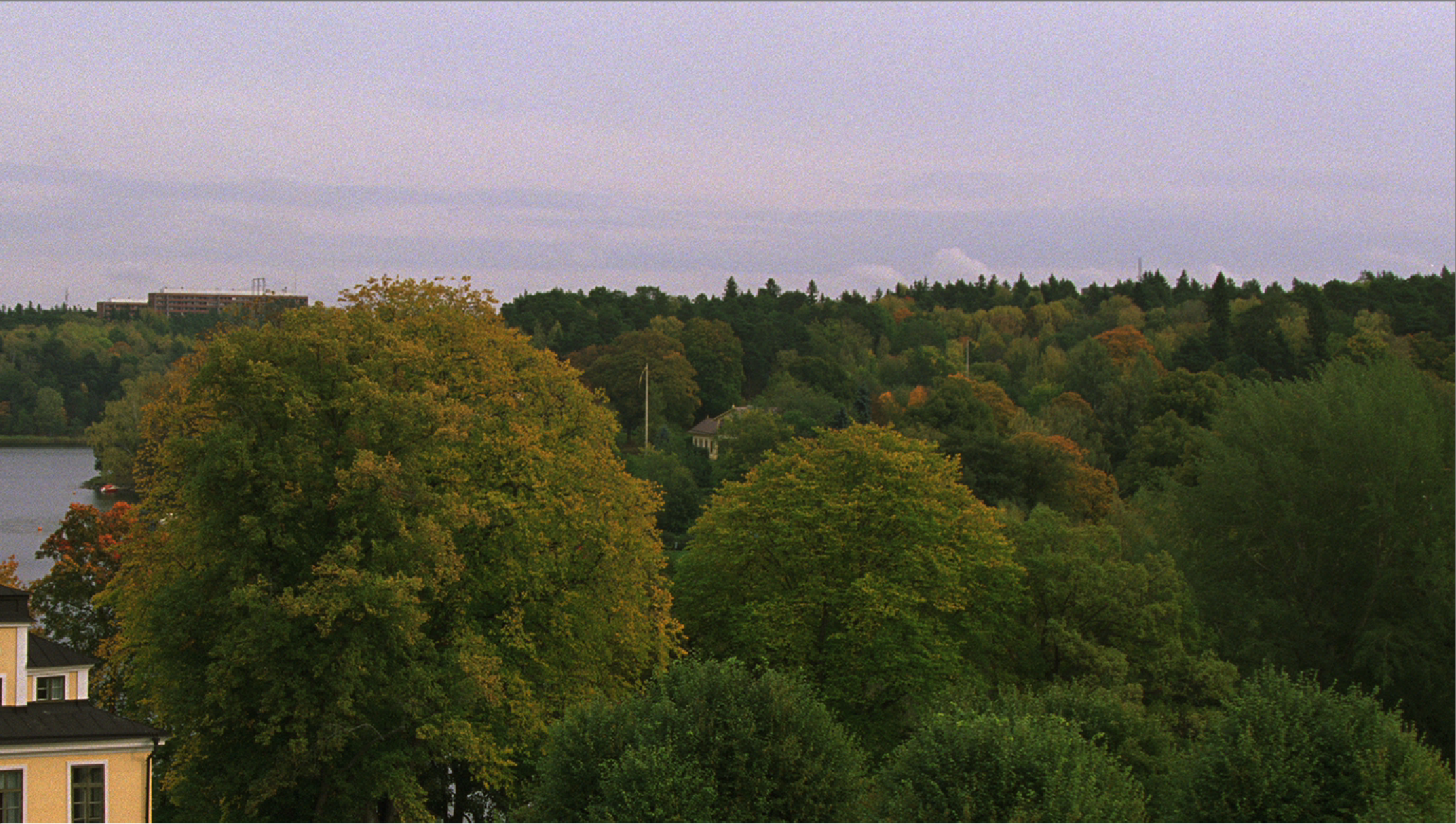}
    \caption{original sequence}
\end{subfigure}
\hfill
\begin{subfigure}{0.45\textwidth}
    \centering
    \includegraphics[width=\textwidth]{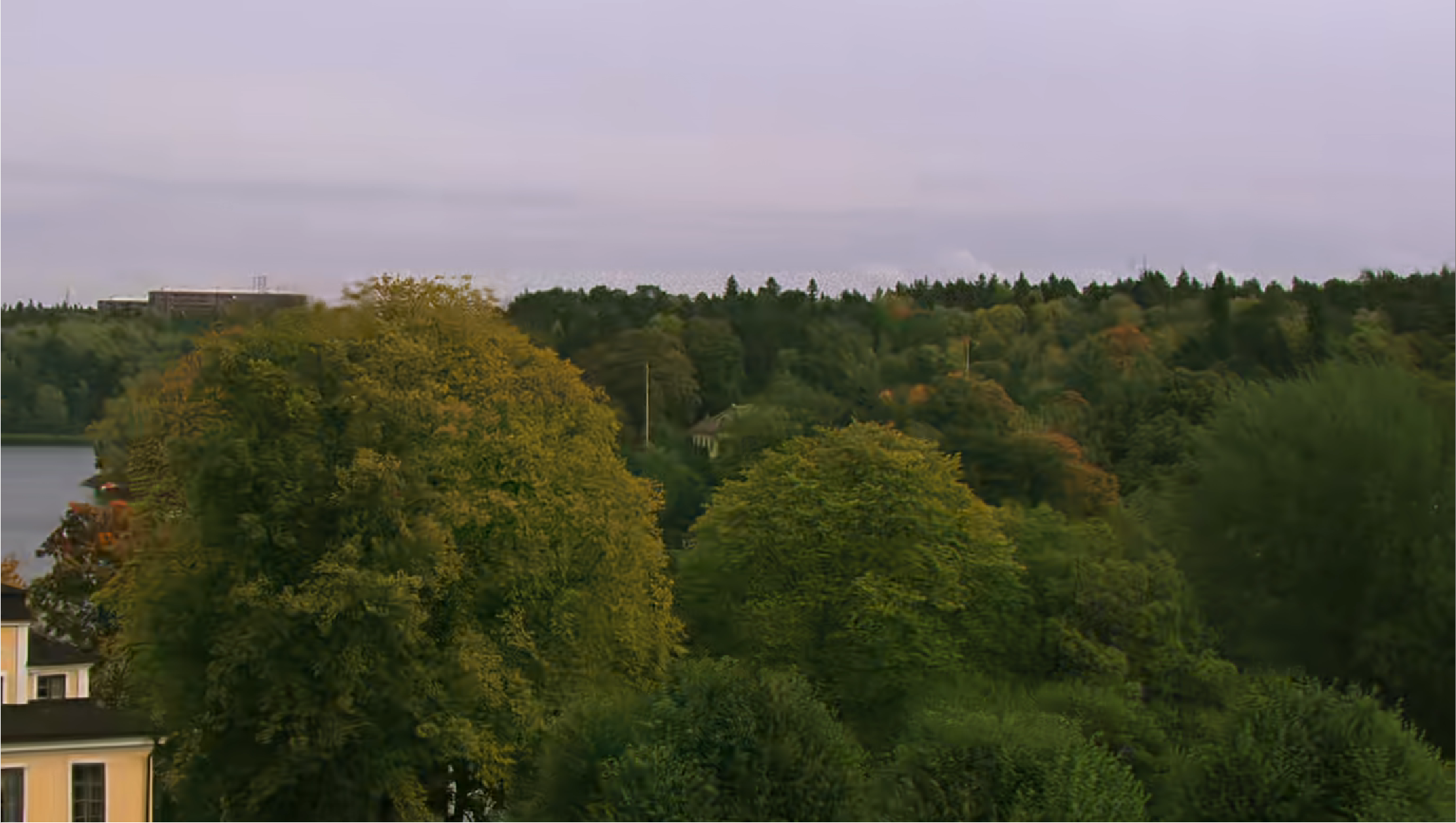}
    \caption{without MCTF}
\end{subfigure}
\hfill
\begin{subfigure}{0.45\textwidth}
    \centering
    \includegraphics[width=\textwidth]{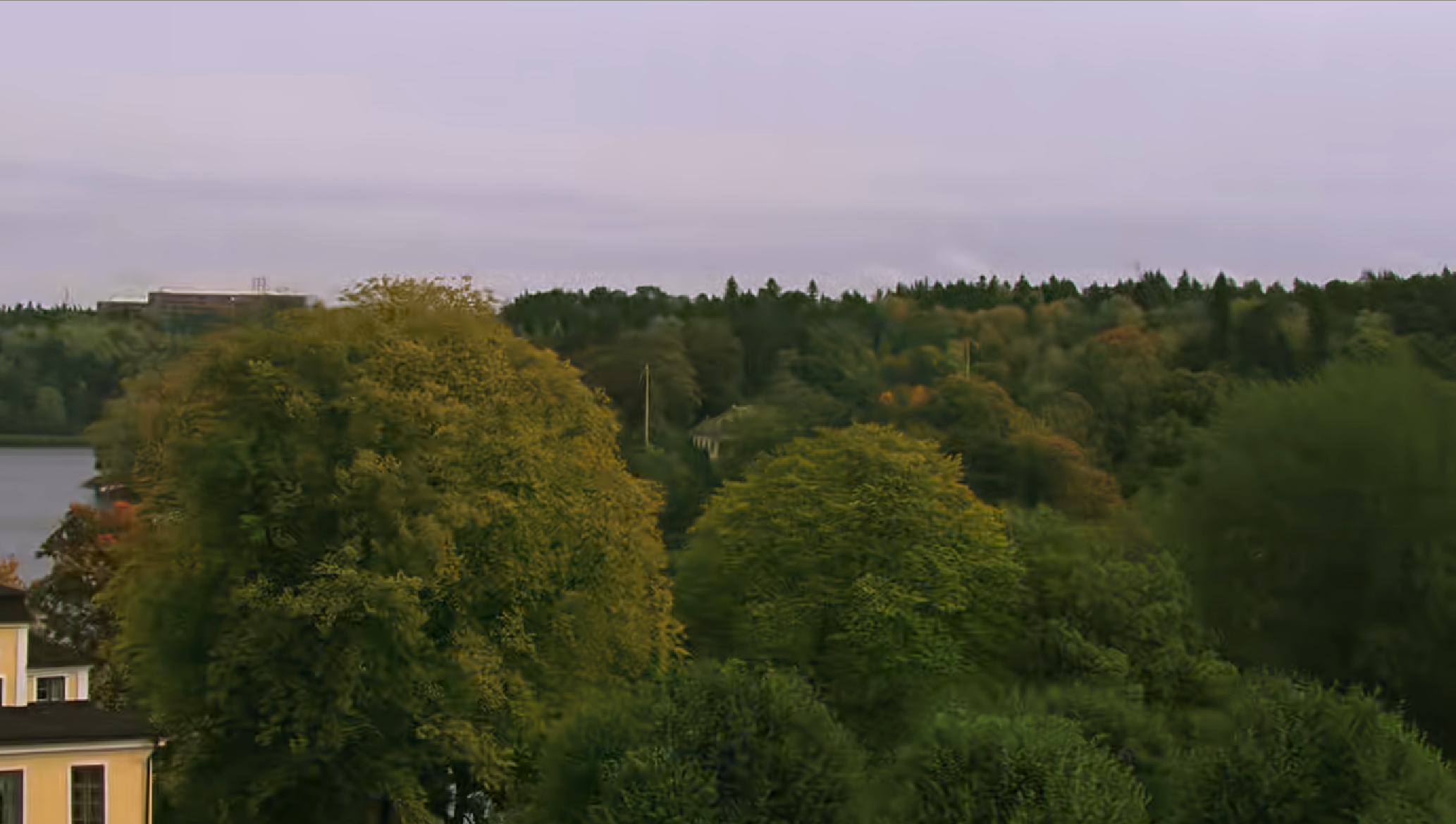}
    \caption{with MCTF}
\end{subfigure}
\hfill
\begin{subfigure}{0.45\textwidth}
    \centering
    \includegraphics[width=\textwidth]{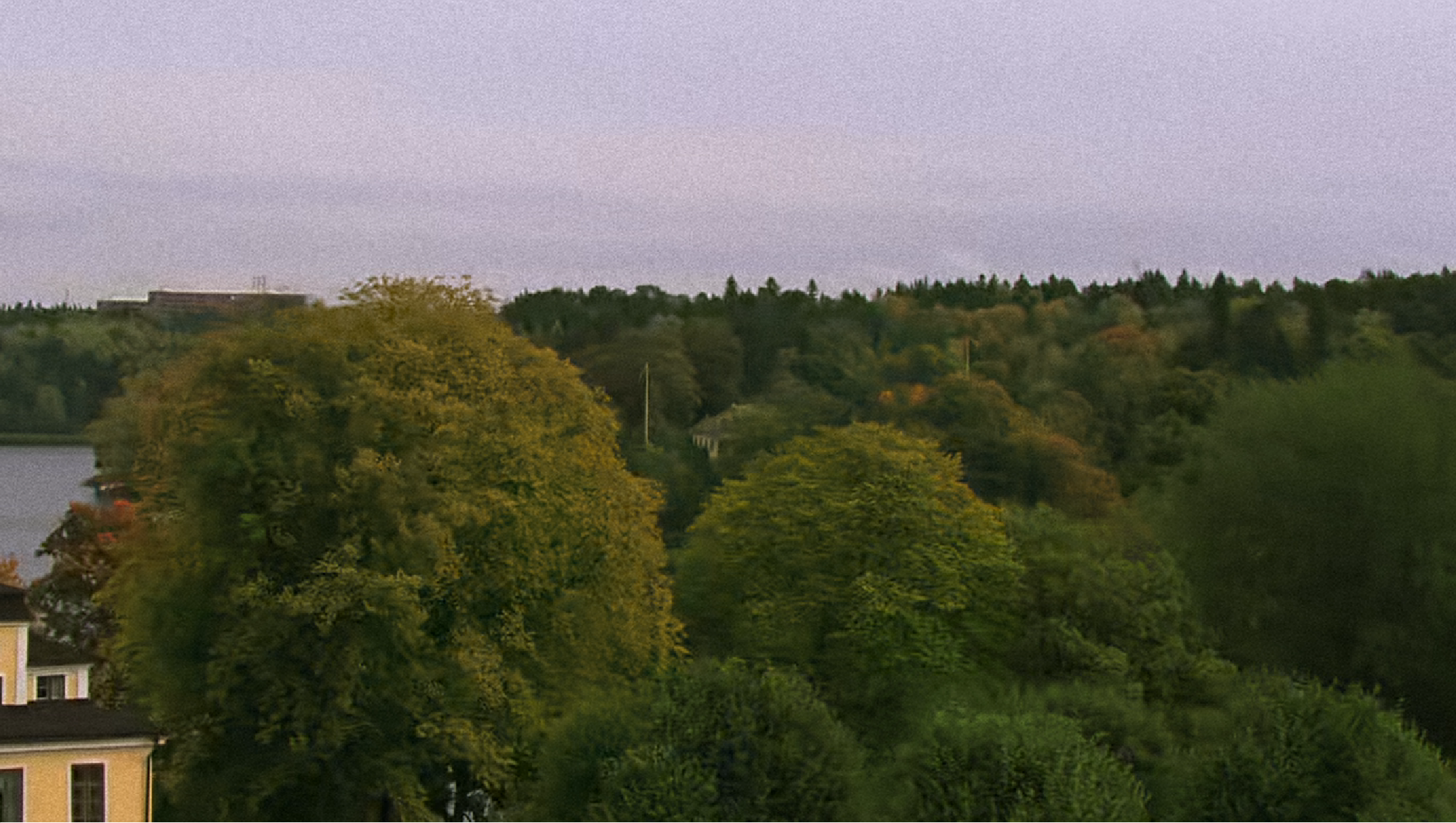}
    \caption{with the proposed toolchain (with FGC)}
\end{subfigure}
\vspace{-0.3em}
\caption{Comparison of the representative (cropped) frame of \textit{IntoTree} test sequence encoded at 0.25\,Mbps.}
\label{fig:intotree_plot}
\end{figure*}

\subsection{Subjective quality analysis}
In scenarios where data is limited, \ie low bitrate encoding, achieving optimal visual quality becomes challenging. Compression artifacts, such as blocking and banding, tend to become more pronounced under these constraints, as shown in the examples in Figure~\ref{fig:ghost_intro_plot}, Figure~\ref{fig:oldtowncross_plot}, and Figure~\ref{fig:intotree_plot}. We observe a controlled noise introduced by FGS, mimicking the film grain characteristics of the original video sequence. This synthesized grain serves as a \emph{visual distraction}, effectively camouflaging compression-related imperfections. By strategically incorporating film grain, we observe an enhancement in the perceptual quality of the video, and viewers are less likely to notice artifacts that might otherwise be more visible in the absence of this synthesized grain~\cite{jvet_grain_view}. At higher bitrates, FGS ensures that the reconstructed video maintains a filmic texture, enabling a more \textit{aesthetically pleasing} viewing.

\subsection{Rate-distortion analysis}
\begin{figure}[t]
\centering
    \includegraphics[width=0.235\textwidth]{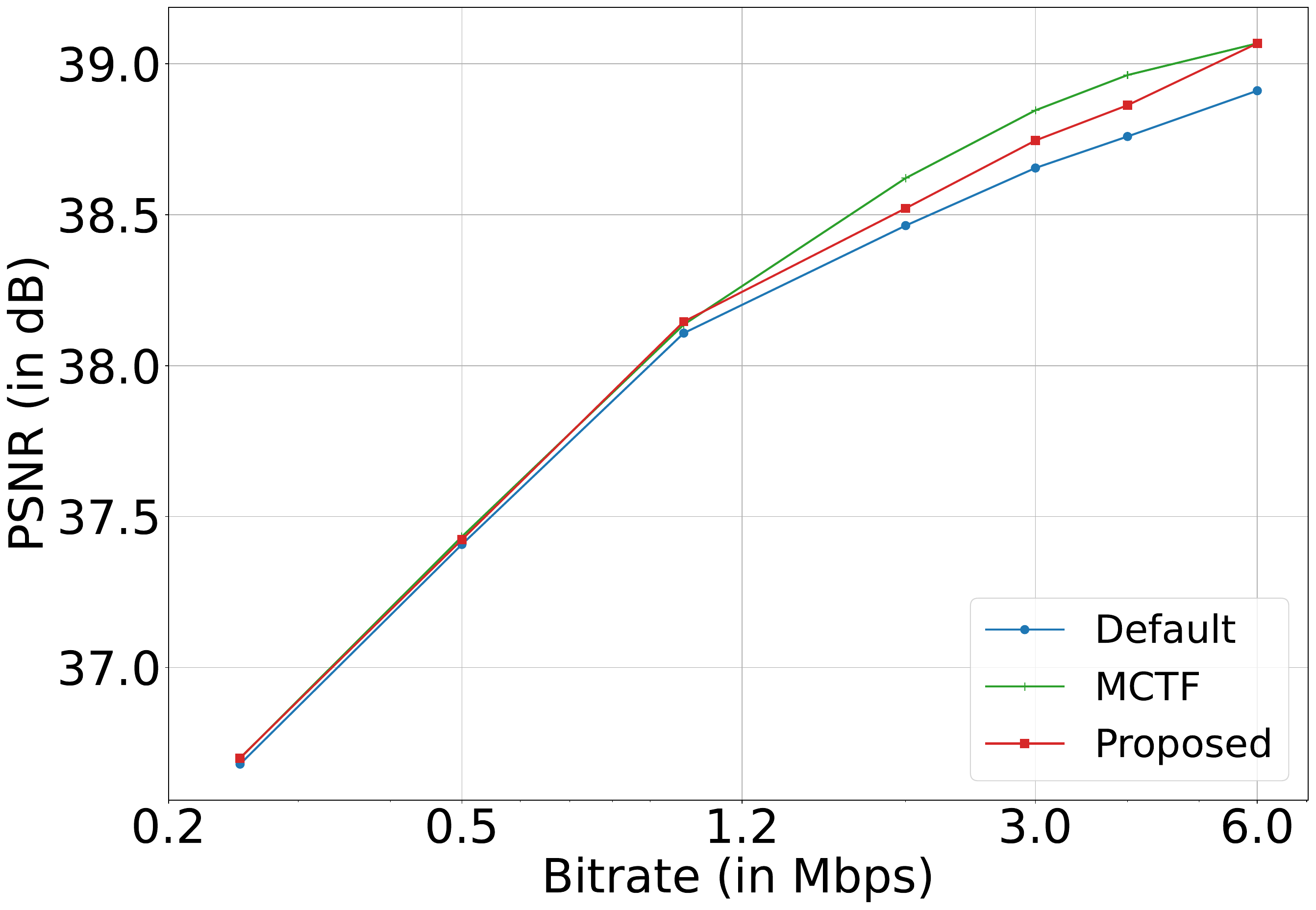}
    \includegraphics[width=0.235\textwidth]{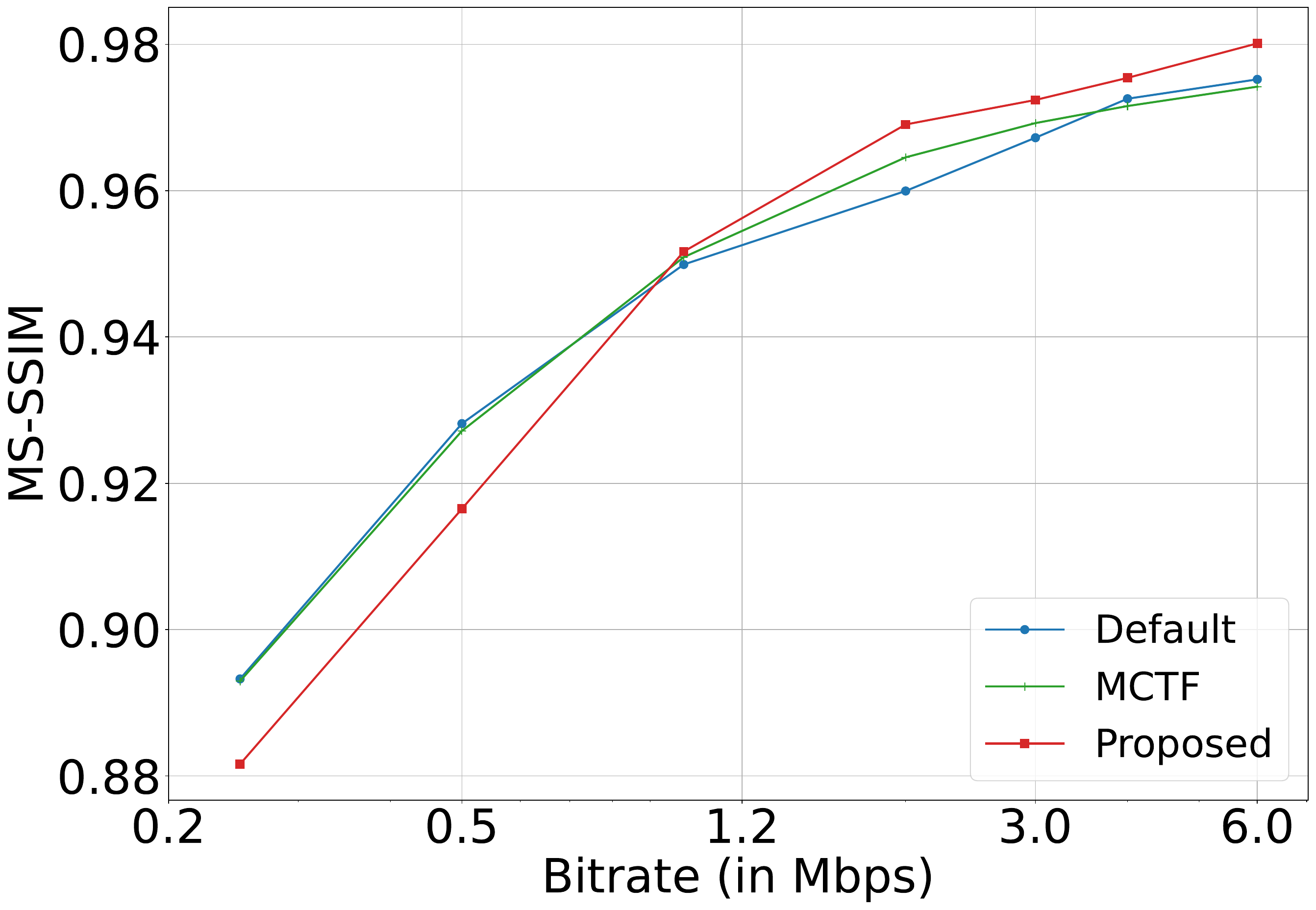}
\caption{RD curves using the \texttt{Default} toolchain, MCTF (without FGS), and the proposed toolchain for \emph{IntoTree} test sequence encoded at \emph{faster} preset.}
\label{fig:intotree_rd}
\end{figure}

Figure~\ref{fig:intotree_rd} analyzes the RD curves of \textit{IntoTree} sequence coded using the \texttt{default} toolchain, MCTF (without FGS), and the proposed toolchain using \emph{faster} preset, and bitrates listed in Table~\ref{tab:exp_par}. 
Since the film grain is filtered out before encoding, the proposed toolchain lowers the bitrate needed to achieve a similar perceptual quality. However, we observe that traditional metrics like PSNR and SSIM are not suitable for evaluating the perceptual quality of film grain coding owing to their lack of texture sensitivity. Furthermore, PSNR and SSIM are sensitive to noise, such that they penalize the addition of film grain, leading to lower scores despite an improvement in perceptual quality. VMAF~\cite{VMAF}, while more advanced, is not trained to evaluate the perceptual quality of VVC-coded videos~\cite{xpsnr_vs_vmaf}.

Given these limitations, specialized metrics focusing on texture enhancement, perception of controlled noise, and overall film-like appearance would be more appropriate for evaluating film grain coding, subject to future work~\cite{sfga_ref}. Metrics that include human perception aspects and consider texture fidelity alongside noise would offer a better assessment of the quality enhancements film grain brings to video content.

\begin{table}[t]
\caption{Runtime complexity of the proposed toolchain compared to the \texttt{default} toolchain with the same preset.}
\centering
\resizebox{0.65\columnwidth}{!}{
\begin{tabular}{l|| c| c| c}
\specialrule{.12em}{.05em}{.05em}
\specialrule{.12em}{.05em}{.05em}
Video & Preset & $\Delta T_{E}$ & $\Delta T_{D}$ \\
& &  [\%] & [\%] \\
\specialrule{.12em}{.05em}{.05em}
\specialrule{.12em}{.05em}{.05em}
IntoTree~\cite{svt_dataset_ref} & \multirow{5}{*}{slower} & -6.92 & 15.32 \\
CampfireParty~\cite{sjtu_ref} &   &  -5.61 & 14.91 \\
OldTownCross~\cite{svt_dataset_ref} &  &  -11.59 & 10.21 \\
CrowdRun~\cite{svt_dataset_ref} &  &  -4.35 & 17.34 \\
ParkScene~\cite{boyce_jvet-j1010_2018}  &  &  -10.44 & 11.24 \\
\hline
IntoTree & \multirow{5}{*}{medium} &  102.47 & 18.77 \\
CampfireParty &   &   91.46 & 18.55\\
OldTownCross &  &  83.71 & 14.68 \\
CrowdRun &  &  97.46 & 18.59 \\
ParkScene  &  &  70.39 & 13.98 \\
\hline
IntoTree & \multirow{5}{*}{faster} &  579.82 & 16.02 \\
CampfireParty &    &  605.03 & 16.92 \\
OldTownCross &   &  678.31 & 19.76\\
CrowdRun &   &  705.35 & 20.20 \\
ParkScene  &   &  583.27 & 16.27 \\
\specialrule{.12em}{.05em}{.05em}
\specialrule{.12em}{.05em}{.05em}
\end{tabular}
}
\label{tab:res_cons}
\end{table}

\subsection{Encoding and decoding time}
This section discusses the encoding and decoding time using the proposed toolchain. This is particularly relevant in adaptive streaming applications, where content adapts to varying network conditions, and handling film grain efficiently within bandwidth constraints is crucial~\cite{DASH_Survey}.

\paragraph{Encoding time} represents the time the encoding server takes to encode the raw input video to the bitstream. Notably, this measure in the proposed toolchain also includes the time taken to estimate film grain. Relative encoding time differences between the bitstreams encoded in the proposed toolchain and the bitstreams encoded in the default toolchain are evaluated as follows:
\begin{equation}
    \Delta T = \frac{\sum \hat{t}_{\text{prop}}}{\sum \hat{t}_{\text{def}}} - 1.
\end{equation}
where $\sum \hat{t}_{\text{def}}$ and $\sum \hat{t}_{\text{prop}}$ represent the sum of encoding times of all bitstreams encoded in the default toolchain and the proposed toolchain, respectively. 

As Table~\ref{tab:res_cons} shows, FGA contributes to the increased relative duration required for encoding as the preset progresses towards \emph{faster} configuration. This increase is also attributed to MCTF being applied on all frames to calculate film grain characteristics in each frame. However, the overall encoding time using the proposed toolchain reduces up to \SI{11.59}{\percent} using \emph{slower} preset. This is attributed to the reduced video content complexity after denoising, making the motion estimation process more accurate and straightforward, offsetting the expected time increments induced by FGA.

\paragraph{Decoding time} represents the time the client device takes to decode the bitstream into the raw video format. However, this measure in the proposed toolchain also includes the time taken by the film grain synthesis phase. The decoding time has increased due to the supplementary computational load of FGS. Furthermore, the effect on decoding time is intertwined with the encoding presets; as encoding presets progress towards slower configurations, the proportional increase in decoding time tends to diminish. This is because slower presets typically yield enhanced compression efficiency, partially counterbalancing the potential increase in decoding time.

\subsection{Adaptive streaming applications}
Integrating film grain analysis (FGA) within the encoding process for adaptive streaming has various advantages, notably optimizing computational resources and preserving cinematic quality across multiple representations. Film grain characteristics, once analyzed, remain consistent across all representations~\cite{jtps_ref} sharing the exact resolution. Leveraging the FGC SEI across these representations facilitates the reuse of the analyzed data, eliminating the need for redundant FGA executions~\cite{emes_ref}. This strategic reuse reduces computational overhead and ensures coherence in the film grain representation across multiple bitrates or qualities. While it is true that different representations may have varying compression artifacts, film grain is an intrinsic part of the content that transcends mere artifacts. Notably, the argument for reusing film grain analysis does not dismiss the presence of compression artifacts but emphasizes the need for consistency in the film grain pattern, which is an intentional and artistic element. Moreover, adjusting the film grain analysis for each representation may introduce inconsistencies, disrupting the intended visual experience and potentially leading to jarring transitions between representations.

\section{Conclusions and future directions}
This paper presented an overview of the film grain handling toolchain for VVC-based open implementation using the VVenC encoder and VTM decoder. The experimental results show that the proposed toolchain improves the subjective quality of the grainy video content encoded at multiple bitrates, compared to \emph{default} toolchain, considering VVenC encoding. Moreover,  using \emph{slower} preset, the proposed toolchain reduces the encoding duration by up to \SI{11.59}{\percent} while the decoding time increases by up to \SI{15.32}{\percent}. These outcomes underscore the inherent trade-offs in optimizing the film grain handling toolchain, emphasizing the need for a balanced approach that prioritizes encoding efficiency and video quality.

In the future, denoising and film grain estimation shall be tuned for various encoding presets in VVenC. Furthermore, more sophisticated models shall be investigated to represent film grain accurately in the digital domain. This could involve advanced statistical models and/or machine learning approaches to capture the intricate characteristics of film grain. Another field of future work is developing quality assessment metrics that better capture human perception of film grain.

\balance
\bibliographystyle{IEEEtran}
\bibliography{references.bib}
\balance
\end{document}